\documentclass[graybox]{svmult}

\usepackage{mathptmx}       
\usepackage{helvet}         
\usepackage{courier}        
\usepackage{type1cm}        
\usepackage{makeidx}         
\usepackage{graphicx}        
\usepackage{multicol}        
\usepackage[bottom]{footmisc}

\makeindex             

\begin{document}

\title*{Classical dynamics of a two-species Bose-Einstein condensate 
in the presence of nonlinear maser processes}
\author{B.~M. Rodr\'{\i}guez-Lara and R.-K. Lee}
\institute{B.~M. Rodr\'{\i}guez-Lara \at Centre for Quantum Technologies, National University of Singapore, Singapore 117542, \email{bmlara@nus.edu.sg}
\and R.-K. Lee \at Institute of Photonics Technologies and Department of Physics, National Tsing-Hua University, Hsinchu 300, Taiwan, \email{rklee@ee.nthu.edu.tw}}
%
%
\maketitle

\abstract{The stability analysis of a generalized Dicke model, in the
semi-classical limit, describing the interaction of a two-species
Bose-Einstein condensate driven by a quantized field in the presence
of Kerr and spontaneous parametric processes is presented. The
transitions from Rabi to Josephson dynamics are identified depending
on the relative value of the involved parameters. Symmetry-breaking dynamics are shown for
both types of coherent oscillations due to the quantized field and
nonlinear optical processes.}

\section{Introduction}
\label{sec:1}
A great deal of attention has been granted to research about unifying concepts in classical and quantum physics through experimental demonstrations involving Bose-Einstein Condensates (BECs) at a macroscopic scale \cite{BEC-book}.
For example, by considering BECs loaded in a double-well potential, the quantum tunnelling between two trapped condensates provides a possibility to study and understand symmetry-breaking, self-trapping, and Josephson oscillation; all of them are fundamental problems in quantum physics.
Macroscopic quantum self-trapping \cite{Albiez} and bosonic Josephson junction \cite{Levy} have been demonstrated recently, and successfully described by a mean-field approach.
The realization of these macroscopic quantum self-trapped modes opens up new avenues for research; \textit{e.g.}, generation of squeezed atomic states \cite{squeeze} and  atomic interferometry \cite{interferometry}.

By using the symmetric and asymmetric stationary eigenstates of the Gross-Pitaevskii equation for a macroscopic condensate trapped in a symmetric double-well potential, the nonlinear Hamiltonian (in units of $\hbar$) describing coherent atomic tunnelling between two zero-temperature BECs is equivalent to that of a nonrigid pendulum \cite{Smerzi, Raghavan}, {\textit i.e.}, 
\begin{eqnarray}
H_{SDW} =  \frac{\Lambda}{2}~z^2 - \sqrt{1-z^2} ~\cos \Phi, \label{eq:HDoubleWell}
\end{eqnarray}
where the length of pendulum decreases with the angular momentum $z$, which denotes the fractional relative population between the condensates in the two wells.
The ratio of the on-site interaction energy and the coupling matrix element is characterized by the parameter  $\Lambda$, while the tilt angle $\Phi$ shows the phase difference between the two condensates.

\begin{figure}[b]
\includegraphics[scale=1]{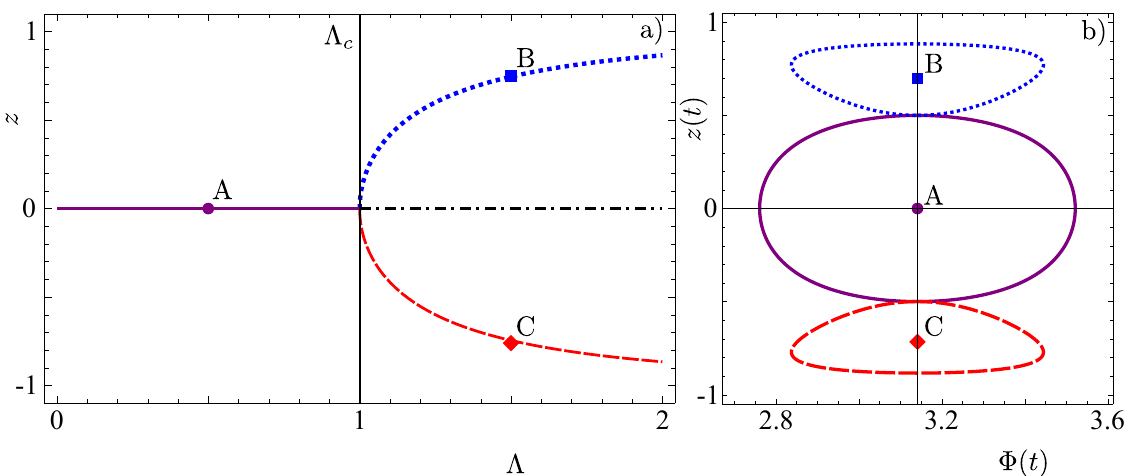}
\caption{ (Color online) a) Fixed points for the equations of motion of a BEC in a double-well potential, described by Eqs.(\ref{eq:DWEqMz}-\ref{eq:DWEqMPhi}), at $\Phi = \pi$. The bifurcation of a single stable fixed point in the Rabi regime (solid purple) split into two fixed points in the Josephson regime (dotted blue and dashed red) is shown at the critical value, $\Lambda_{c}=1$. The unstable fixed point, or separatrix, in the Josephson regime (dash-dotted black) is also shown. b) Examples of trajectories for identical Rabi oscillations (solid purple) around the fixed point A ($\Lambda=0.5$) and localized Josephson oscillations around the fixed points B (dotted blue ) and C (dashed red ) ($\Lambda=1.5$) are demonstrated with symmetrical initial conditions $\{z_{\pm}=\pm 0.5, \Phi=\pi\}$.}
\label{fig:Fig1}       
\end{figure}

This nonlinear system allows for macroscopic quantum self-trapping (\textit{i.e.}, localized oscillations of the fractional population difference $z$), which can be easily deduced from the equations of motion of the system, 
\begin{eqnarray}
\dot{z} &=& - \sqrt{1 - z^2}~ \sin \Phi, \label{eq:DWEqMz}  \\
\dot{\Phi} &=&  \Lambda z + \frac{z}{\sqrt{1-z^2}}~\cos \Phi. \label{eq:DWEqMPhi}
\end{eqnarray}
The fixed points of such equations of motion present a pitchfork bifurcation at the critical point $\Lambda_{c} = 1$; \textit{i.e.}, $\{z_{\pi}=0, \Phi=\pi\}$ for $\Lambda < 1$ and $\{z_{\pi}= \pm \sqrt{\Lambda^2 -1} / \Lambda, \Phi=\pi\}$ for $\Lambda > 1$.
This bifurcation signals a symmetry-breaking in the dynamics of the system from Rabi to Josephson dynamics; such that  for symmetric initial conditions, $\{z_{\pm} = \pm z_{0}, \Phi = \pi\}$, the trajectories will be identical in the Rabi regime, $\Lambda < 1$, but localize in opposite hemispheres of phase space in the Josephson regime, $\Lambda > 1$.
Figure \ref{fig:Fig1} shows an example of such symmetry-breaking phenomena.
The transition from Rabi to Josephson dynamics due to the pitchfork bifurcation of the fixed points has been confirmed experimentally in an equivalent model consisting of two hyperfine states of a
single atomic specie BEC coupled by a classical two-photon transition in the semi-classical limit \cite{Zibold}. 

Recently, strong coupling between a BEC and the quantized field mode of an ultrahigh-finesse optical cavity has been demonstrated \cite{BEC-cavity-1, BEC-cavity-2, BEC-cavity}.
The spectra of these strongly-coupled systems shows a level splitting attributed to different hyperfine structures of the given atomic species.
Motivated by these experiments, here, we consider a gaseous BEC, composed of bosonic atoms populating two hyperfine levels, coupled to a quantized field cavity mode, including nonlinear interactions among the ultracold atoms.
Instead of two trapped BECs interacting through quantum tunnelling in a double-well potential, the hyperfine levels of the two-species condensate are driven by the quantized cavity field.
It will be demonstrated later that the Hamiltonian describing this system consisting of a BEC in a cavity  can be reduced into a nonlinear Dicke model with an additional atom-atom quadratic interaction term; in the semi-classical limit, this system can be though of as a pendulum with changeable mass.
In order to study a general model, we consider second and third order nonlinear processes for the field.
Our goal is to present a steady-state analysis of the equations of motion in the large-ensemble-size limit for the collective dynamics of this system.
In Section \ref{sec:2}, the model and possible physical realizations are presented and discussed.
In order to present analytical results, a weak regime is defined in Section \ref{sec:3}  for weak coupling and nonlinearities (compared to the frequency of the driving field). 
The symmetry of Josephson dynamics is shown to break by the driving quantum field.
Finally, Section \ref{sec:4} closes with a summary.

\section{Model}
\label{sec:2}
The proposed model Hamiltonian describing the interaction of a single electromagnetic cavity mode $\hat{a}$ coupled to a BEC with two internal hyperfine structure levels $\hat{b}_{\downarrow}$ and $\hat{b}_{\uparrow}$ can be obtained from the Gross-Pitaevskii equation describing a two-species BEC interacting with a quantum field~\cite{EPL} and is written as,
\begin{eqnarray}
\hat{H}_0 &=& \omega_0 \hat{a}^\dagger \hat{a} + \sum_{j=\uparrow, \downarrow}(E_j \hat{b}_j^\dagger \hat{b}_j + \frac{1}{2}G_{jj}\hat{b}_j^\dagger \hat{b}_j^\dagger \hat{b}_j \hat{b}_j) \nonumber\\
&& + \frac{g}{\sqrt{N}} \left(\hat{a} + \hat{a}^\dagger \right) \left(\hat{b}_\uparrow^\dagger \hat{b}_\downarrow+ \hat{b}_\downarrow^\dagger \hat{b}_\uparrow \right) + G_{\uparrow\downarrow}\hat{b}_\uparrow^\dagger \hat{b}_\downarrow^\dagger \hat{b}_\downarrow\hat{b}_\uparrow, 
\label{eq-H}
\end{eqnarray}
where the frequency $\omega_0$ is the cavity mode frequency, the parameter $g$ is the coupling strength between the cavity mode and condensed atoms, and $N$ is the number of atoms in the condensate.
The energies of two internal hyperfine levels are labelled as $E_\downarrow$ and $E_\uparrow$ with an intra-atomic transition frequency $\omega_a \equiv E_\uparrow - E_\downarrow$. Here, we suppose the interaction Hamiltonian as a coupled two-component BEC \cite{CLee-2}, by introducing $G_{jj}$ ($j=\uparrow$ and $\downarrow$) and $G_{\uparrow\downarrow}$ for the inter-atomic and intra-atomic interactions, respectively.

The Hamiltonian in Eq.(\ref{eq-H}) can be further reduced by regarding all atoms as spin $1/2$ particles and defining collective angular momentum operators via Schwinger transformation,  
$\hat{J}_{x}=(\hat{b}_{\uparrow}^{+}\hat{b}_{\downarrow}+\hat{b}_{\uparrow}\hat{b}_{\downarrow}^{+})/ 2 $, $\hat{J}_{y}=(\hat{b}_{\uparrow}^{+}\hat{b}_{\downarrow}-\hat{b}_{\uparrow}\hat{b}_{\downarrow}^{+}) / 2i$, and $\hat{J}_{z}=(\hat{b}_{\uparrow}^{+}\hat{b}_{\uparrow}-\hat{b}_{\downarrow}^{+}\hat{b}_{\downarrow})/ 2$, alongside the raising and lowering operators $\hat{J}_{+}=\hat{b}_{\uparrow}^{+}\hat{b}_{\downarrow}$ and  
$\hat{J}_{-}=\hat{b}_{\uparrow}\hat{b}_{\downarrow}^{+}$.
The reduced Hamiltonian is given by the expression, 
\begin{eqnarray}
\hat{H}_{DLMG}=\Delta \hat{J}_z + \frac{g}{\sqrt{N}}\left(\hat{a} + \hat{a}^\dagger \right) \hat{J}_{x} +  \frac{\xi}{N} \hat{J}_z^2,
\label{eq-nlin-Dicke}
\end{eqnarray}
where we have dropped the constant energy term 
$\frac{N}{2}[E_{\uparrow}+E_{\downarrow}-\frac{1}{2}(G_{\uparrow\uparrow}+G_{\downarrow\downarrow})]+\frac{N^{2}}{8}(G_{\uparrow\uparrow}+G_{\downarrow\downarrow}+2 G_{\uparrow\downarrow})$.
The average interaction energy of each atom is defined as $\xi/N = \frac{1}{2}(G_{\uparrow\uparrow}+G_{\downarrow\downarrow}-2G_{\uparrow\downarrow})$ to account for the collective interaction among the condensates. The frequency detuning between cavity field and hyperfine transition is given by $\Delta = \omega_a-\omega_0$. 
In the following analysis, only positive values of $\xi$ are considered (repulsive interaction) for the case of intra-species interaction, $G_{\uparrow\uparrow}$ or $G_{\downarrow\downarrow}$, larger than the inter-species interaction, $G_{\uparrow\downarrow}$.

Equation (\ref{eq-nlin-Dicke}), the starting Hamiltonian for our analysis, corresponds to a generalized Dicke Hamiltonian without the rotating wave approximation and an additional atom-atom quadratic nonlinear  interaction.
It is well known that the Dicke Hamiltonian describes the collective dynamics for an ensemble of two-level systems driven by a quantum field cavity mode within a quantum electrodynamics (QED) configuration \cite{Dicke}.
From many-body physics, the Hamiltonian in Eq. (5) is also equivalent to the Lipkin-Meshkov-Glick (LMG) model in the limit $g = 0$ \cite{Lipkin1965p188}. 
Through the interaction among ensemble atoms, the LMG model, originally for $N$ fermions distributed in two $N$-fold degenerate levels and interacting via a monopole-monopole force, was used to describe the Josephson effect in a two-species BEC and found to produce maximal pairwise entanglement of formation at the phase transition of its ground state \cite{Vidal2004p022107, Vidal2004p062304}. 
Schemes proposed to implement a dissipative LMG model in optical cavity-QED~\cite{Morrison2008p040403} and in circuit-QED~\cite{Larson2010p54001} have been discussed. 
The Hamiltonian in Eq. (5) hereby  will be called the Dicke-LMG (DLMG) Hamiltonian.
Recently, it has been theorized that the ground state of the DLMG model supports phase transitions~\cite{Chen2010p053841}.
In the semi-classical limit, both pitchfork and asymmetric bifurcations of the stable fixed points were found for this model \cite{PRE}.
Also, the full quantum analysis of such a system showed a highly entangled ground state near the values of the semi-classical critical parameters \cite{RodriguezLara2010}.

Now, we extend the DLMG Hamiltonian to include a Kerr medium \cite{Chi, Deb} and degenerate parametric amplification \cite{Collett} with nonlinear parameters $\kappa$ and $\chi$, respectively. The model Hamiltonian for such a system is given by the expression:
\begin{eqnarray} \label{eq:FullHamiltonian}
\hat{H} = \omega_{0} \hat{a}^{\dagger} \hat{a} + \kappa ( \hat{a}^{\dagger} \hat{a} )^2 + \chi \left(\hat{a}^{2} + \hat{a}^{\dagger 2}\right) + \omega_{a} \hat{J}_{z} + \frac{\xi}{N} \hat{J}_{z}^{2} + \frac{g}{\sqrt{N}} \left( \hat{a} + \hat{a}^{\dagger} \right) \hat{J}_{x}, 
\end{eqnarray}
where the modified field frequency $\omega_{0} = \omega_{f} - \kappa + 2 \chi$ involves the modifications arising from the original Kerr and parametric amplifier terms, $\hat{a}^{\dagger 2}\hat{a}^{2}$ and $\left( \hat{a} + \hat{a}^{\dagger} \right)^2$. 
Experimental realizations providing an assorted range of tunable parameters for the DLMG model may include a two-hyperfine-structure-defined-modes BEC coupled to a quantum cavity field mode through a one microwave photon process; \textit{e.g.}, trapped hyperfine ground states of a Sodium BEC inside a microwave cavity~\cite{Gorlitz}. 
Arrays of interacting superconducting qubits coupled to the quantum field mode of a coplanar waveguide resonator may be considered as a physical realization limited by small ensemble sizes~\cite{Tsomokos2010}. 

In the literature, similar model Hamiltonians have been considered without the feedback from the BEC on the electromagnetic field; \textit{e.g.}, a second-order phase transition from immiscible to miscible is revealed in such a two-species systems by considering a linear mixing between the binary components~\cite{miscibility}.
Stable domain-wall solutions, on top of flat continuous wave asymmetric bimodal states, can also be found near the point of the symmetry-breaking bifurcations ~\cite{domainwalls}.
The relative phase of domain-walls and breather-like dynamics for these dressed two-species BECs has also been studied by considering a classical external driving field~\cite{domain-em, dressed}. 
Our results with a driven quantum field differs considerably from earlier studies:
It is found that even in the semi-classical limit, the quantum field drive manifests in the excitation ratio parameter.
When this excitation parameter is small or close to one, a discrete total excitation modifies the classical dynamics of the system strongly, producing localized asymmetric dynamics where some of the phase space trajectories present a running phase; \textit{e.g.}, Fig.\ref{fig:Fig3}a and Fig.\ref{fig:Fig3}b.  

The proposed Hamiltonian in Eq. (6) may be experimentally realized by a two-species BEC, where the species are defined by two hyperfine-structure levels.
The condensate is confined and  coupled to a quantum cavity field mode through a one-microwave-photon process in the presence of a Kerr and a $\chi^{(2)}$ medium. 
In general, the model described by Hamiltonian (\ref{eq:FullHamiltonian}) has complex dynamics that deserve a study on their own. 
In order to give a clear interpretation, here, we limit ourselves to an analytical approach in the weak regime where intra-ensemble and ensemble-field couplings, as well as nonlinearities, are small compared to the field and hyperfine transition frequencies.

\section{Weak regime}
\label{sec:3}
The full Hamiltonian in Eq. (\ref{eq:FullHamiltonian}) does not conserve the total number of excitations, $\hat{\mathcal{N}} = \hat{a}^{\dagger} \hat{a} + \hat{J}_{z}$, \textit{i.e.}, $[\hat{H}, \hat{\mathcal{N}}] \neq 0$. 
This fact prevents a simple approach like those given in the literature \cite{Zibold, PRE}.
This inconvenience may be bridged in the regime where nonlinearities and couplings are weak compared to the field and transition frequencies.
In this weak regime it is possible to define a couple of unitary transformations \cite{JOSAB}
\begin{eqnarray}
\hat{T} &=& e^{\eta \left(\hat{a}^{2} - \hat{a}^{\dagger 2}\right) },\\
\hat{U} &=& e^{-i \nu \left(\hat{a} + \hat{a}^{\dagger}\right) \hat{J}_{y} }.
\end{eqnarray}
The first of these transformations is equivalent to consider a squeezed basis; while the second one is similar to a polariton transformation and provides an effective rotating wave approximation, with the introduction of the small parameters $\eta = \chi/\omega_{0} \ll 1$ and $\nu = g(1 - 2 \eta)/\left(\omega_a + \omega_{0} - 4 \chi \eta\right) \ll 1$, respectively. 
The following effective Hamiltonian, up to a constant and in units of $\hbar$, is obtained by neglecting all the products of couplings with nonlinearities (\textit{i.e.}, $\kappa$, $\xi$, $\chi$,  $\lambda$, $\eta$ and $\nu$ are all at least a couple orders of magnitude smaller than the field and atom frequencies) and moving into the frame defined by the total excitation number rotating at frequency $\omega = \omega_{0} - 4 \chi \eta$,
\begin{eqnarray} \label{eq:WeakHamiltonian}
\hat{H}_{eff} = \delta \hat{J}_{z} + \kappa \left( \hat{a}^{\dagger} \hat{a} \right)^{2} + \frac{\xi}{N} \hat{J}_{z}^{2} + \frac{\lambda}{\sqrt{N}} \left( \hat{a} \hat{J}_{+} + \hat{a}^{\dagger} \hat{J}_{-} \right).
\end{eqnarray}
Here, the frequency detuning is given by $\delta = \omega_{a} - \omega$ and the effective ensemble-field coupling is $\lambda = \omega \nu$.
This effective Hamiltonian in the weak regime is nothing else than the extended Dicke model studied in reference \cite{RodriguezLara2010} plus a Kerr term.
The Hamiltonian in Eq. (9) conserves the total number of excitations, $[\hat{H}_{eff}, \hat{\mathcal{N}}]=0$.
Note that the exact dynamics defined by Hamiltonian Eq.~(\ref{eq:WeakHamiltonian}) can be calculated by quantum inverse methods \cite{Bogoliubov}.

\subsection{Semi-classical limit}
\label{sec:3.1}
In order to  link our proposed configuration to a generalized pendulum problem, we apply the mean-field approach to study the semi-classical dynamics of this two-species BEC coupled to a optical cavity mode.
It is possible to approximate the expectation values by considering the system in a separable state composed of a coherent photon state \cite{Sudarshan, Glauber}, $\vert \sqrt{n} e^{\imath \phi} \rangle$, and a coherent spin $1/2$ state \cite{Nienhuis, Arecchi}, $\vert z , \theta \rangle$, respectively.
The expectation values for the field with the photon number $n$ and optical phase $\phi$  are:
\begin{eqnarray}
\langle \hat{a} \rangle &\equiv & \alpha = \sqrt{n} ~e^{i \phi}, \\
\langle \hat{a}^{\dagger} \rangle &\equiv& \alpha^{\ast} =  \sqrt{n} ~e^{-i \phi},  \\
\langle \hat{a}^{\dagger} \hat{a} \rangle &\equiv& \vert \alpha \vert^{2} = n .
\end{eqnarray}
If the ensemble is large enough, such that $N \gg 1$ but less than the restriction brought by the two-mode approximation \cite{EPL, Milburn}, the expectation values for the ensemble operators may be approximated by those in the thermodynamic limit, 
\begin{eqnarray}
\langle \hat{J}_{z} \rangle &\equiv& J_{z} \approx  \frac{N}{2} ~z, \\
\langle \hat{J}_{\pm} \rangle &\equiv & J_{\pm} \approx \frac{N}{2} \sqrt{1 - z^2} ~e^{\pm i \theta},
\end{eqnarray}
where the fractional population difference is defined by the rotating angle in the corresponding atomic Bloch sphere, {\textit i.e.}, $z = cos \theta$.
The conserved quantity in our system is the mean total excitation number, which is given by the expression:
\begin{eqnarray}
\langle \hat{\mathcal{N}} \rangle \equiv \mathcal{N} = n + \frac{N}{2} ~z.
\end{eqnarray}
This mean-field  approximation plus the definition of a total phase variable, $\Phi = \phi + \theta$, and an excitation ratio, $k = 2 \mathcal{N}/N$,  allow us to write the weak regime effective Hamiltonian in units of $\hbar N \lambda / 2$ and up to a constant as:
\begin{eqnarray} \label{eq:EffMeanEnergy}
H = \left(\Delta + \frac{\Lambda}{2} z \right) z + \sqrt{2 (k - z)(1 - z^2)} \cos \Phi, 
\end{eqnarray}
where the re-scaled transition detuning now is shifted by the self-phase modulation from the Kerr nonlinearity, $\Delta = (\delta + \kappa)/ \lambda$.
Moreover, the re-scaled coupling ratio is defined as $\Lambda = \xi / \lambda$.
This mean-field Hamiltonian is equivalent with that of the DLMG model in \cite{RodriguezLara2010}, with the difference that in this case the Kerr and $\chi^{(2)}$ nonlinearities play important roles in the frequency detuning $\Delta$ and the characteristic interaction ratio $\Lambda$, respectively.

Equation (16) may be viewed as a more general pendulum with nonlinear pendulum length and changeable pendulum mass described by the excitation ratio, $k$. 
This distinguish our system from the double-well configuration which is equivalent to a pendulum with just nonlinear pendulum length.
Also, note the restriction $k - z > 0$ induced by the model Hamiltonian.
Our system is equivalent to the case of a BEC in an asymmetric double well via a phase $\pi$-shift and a restriction given by $k-z = 1/2$,
\begin{eqnarray}
H_{ADW} = \left( \Delta + \frac{\Lambda}{2}~z\right)~z - \sqrt{1-z^2} ~\cos \Phi, \label{eq:HADoubleWell}
\end{eqnarray}

\subsection{Fixed points of the system}
The dynamics of the semi-classical system is given by the equations of motions for the dimensionless fractional population difference and total phase variable,
\begin{eqnarray} 
\dot{z} &=& \sqrt{2 \left(1- z^{2} \right) \left( k - z \right)} \sin \Phi, \label{eq:WLEqMz}\\ 
\dot{\Phi} &=& \Delta + \Lambda z - \frac{1 + 2 k z + 3 z^2}{\sqrt{2  \left(1- z^{2} \right) \left( k - z \right)}} \cos \Phi . \label{eq:WLEqMPhi} \end{eqnarray}
Stationary states are found at the total phase variables values of $\Phi= 0$ and $\pi$.
Due to the $\pi$ phase difference with Eq. (\ref{eq:HDoubleWell}), here the plasma and $\pi$ oscillations will exchange places appearing at $\Phi=\pi$ and $\Phi=0$, respectively.

The fixed points of this Hamiltonian coincide with the critical points as $\dot{z} \equiv \partial H / \partial \Phi$ and $\dot{\Phi} \equiv \partial H / \partial z$ \cite{PRE}.
Stationary states are found for the phase variable values $\Phi = 0, \pi$ and the excitation parameter value
\begin{eqnarray}\label{eq:ExcParam}
k &=& \frac{3 z^2 - 1}{2 z} + \frac{ (1-z^{2})\vert(\Delta+  \Lambda z) \vert}{4 z^2}\times \left\{ \vert \Delta +  \Lambda z \vert \pm \left[ \left( \Delta +  \Lambda z \right)^2 - 4  z \right]^{1/2} \right\}.
\end{eqnarray}
Notice that, in order to obtain a real excitation ratio, $k$, the allowed fractional population difference is bounded to the range $z \in [-1, z_{-}] \cup [z_{+},1]$, where $z_{\pm} = [ 2 -   \Delta \Lambda \pm  2 (1 - \Delta \Lambda)^{1/2} ] /  \Lambda^2 $ sets the condition $\Delta \le 1 /  \Lambda$.

\begin{figure}[b]
\includegraphics[scale=1]{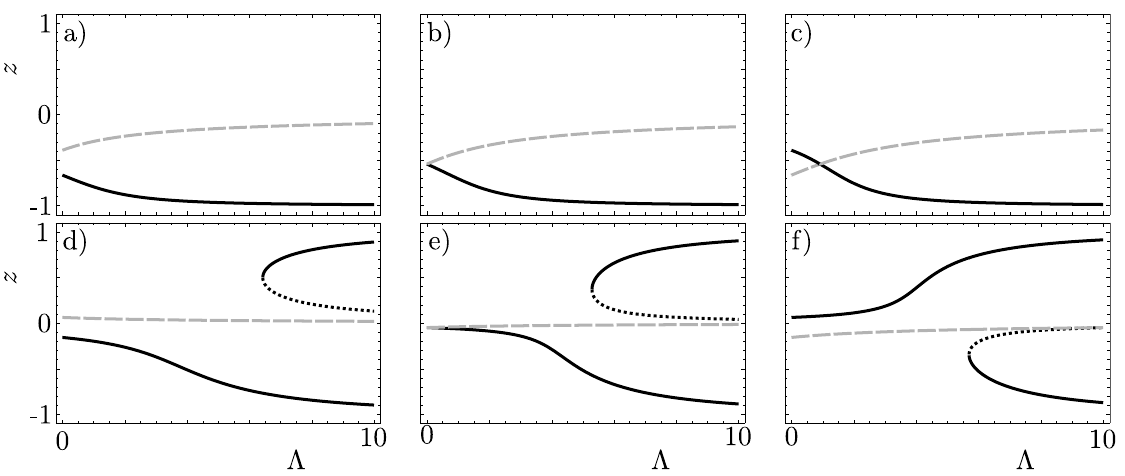}
\caption{ Example of fixed points in the weak regime, described by Eqs.(\ref{eq:WLEqMz}-\ref{eq:WLEqMPhi}), for excitation ratios (a-c) $k=0.1$ and (d-f) $k=10$, respectively. Different tunning ratios, $\Delta = (\delta + \kappa)/\lambda$, are shown: (a,d) $\Delta = -0.5$, (b,e) $\Delta=0$, (c,f) $\Delta=0.5$. Solid and dotted black lines correspond to stable and unstable fixed points for $\pi-$oscillations;  while dashed grey lines correspond to fixed points for plasma oscillations.}
\label{fig:Fig2}       
\end{figure}

An example of fixed points in the weak regime is shown in Fig. \ref{fig:Fig2}.
When the value of the excitation ratio $k$ is less than or within the order of the magnitude of one, we have a quantum drive compared to a classical one ($k \gg 1$).
It is the the introduction of the quantum drive and nonlinear processes that brings a peculiar breaking of the symmetry, different from the pitchfork bifurcation of the classical driving.
It is possible to numerically sample the parameter space, $\{\Delta, \Lambda, k \}$, and see that for any given excitation ratio, $k$, the fixed points satisfy mirror inversion at $\Lambda=0$; \textit{i.e.}, $z(\Lambda, \vert \Delta \vert) = z(-\Lambda,-\vert \Delta \vert)$. 
Also, if the frequency detuning, $\delta$, is set to compensate the Kerr nonlinearity, $\delta = - \kappa$, it is possible to recover results that have been studied in the past; \textit{e.g.}  Fig. \ref{fig:Fig2}(b) and \ref{fig:Fig2}(e) correspond to  Fig. 2 in \cite{PRE}.

\begin{figure}[b]
\sidecaption
\includegraphics[scale=1]{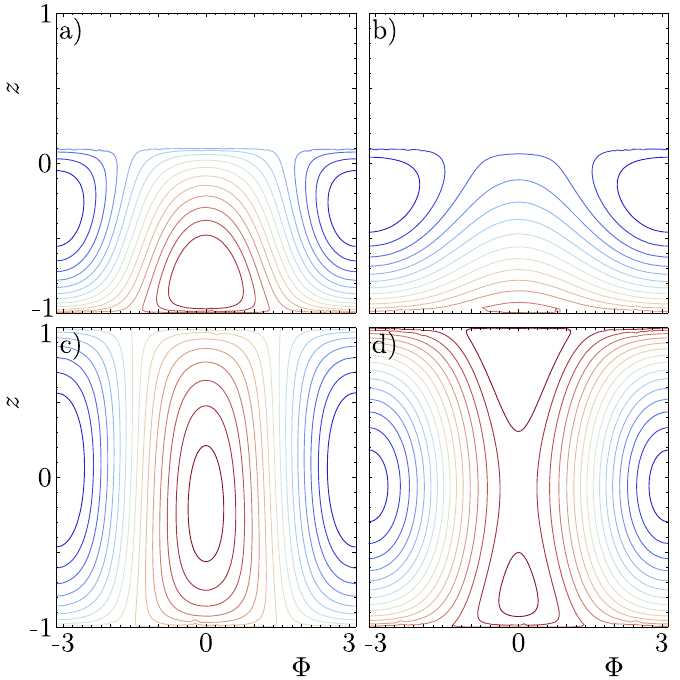}
\caption{ (Color online) Normalized mean value of the effective Hamiltonian energy in Eq. (\ref{eq:EffMeanEnergy}), where a compression of the available phase space is shown for parameter sets $\{ \Delta, \Lambda, k\} =$ (a) $\{-0.5,1,0.1 \}$ and (b) $\{0.5,8,0.1 \}$. Also, the existence of asymmetric trajectories for symmetric initial parameters is intuited from the lack of symmetry with respect to the horizontal axis of the normalized mean value effective energy for parameter sets in the Rabi and Josephson regimes, $\{ \Delta, \Lambda, k\} =$ (c) $\{-0.5,1,10 \}$ and (d) $\{0.5,8,10 \}$, in that order.}
\label{fig:Fig3}       
\end{figure}

From the mean-field Hamiltonian in Eq. (\ref{eq:EffMeanEnergy}), it is straightforward to see that the quantum drive restricts the phase space accessible to Rabi oscillations for low excitation ratio, $k<1$, in order to keep the mean effective energy real, as shown in Fig. \ref{fig:Fig3}(a) and \ref{fig:Fig3}(b).
Figures \ref{fig:Fig3}(c) and \ref{fig:Fig3}(d) allow us to see that the quantum drive also breaks the symmetry of symmetric initial conditions as isoenergetic lines are not symmetric with respect to the horizontal axis when the whole phase space is accessible.

\begin{figure}
\includegraphics[scale=1]{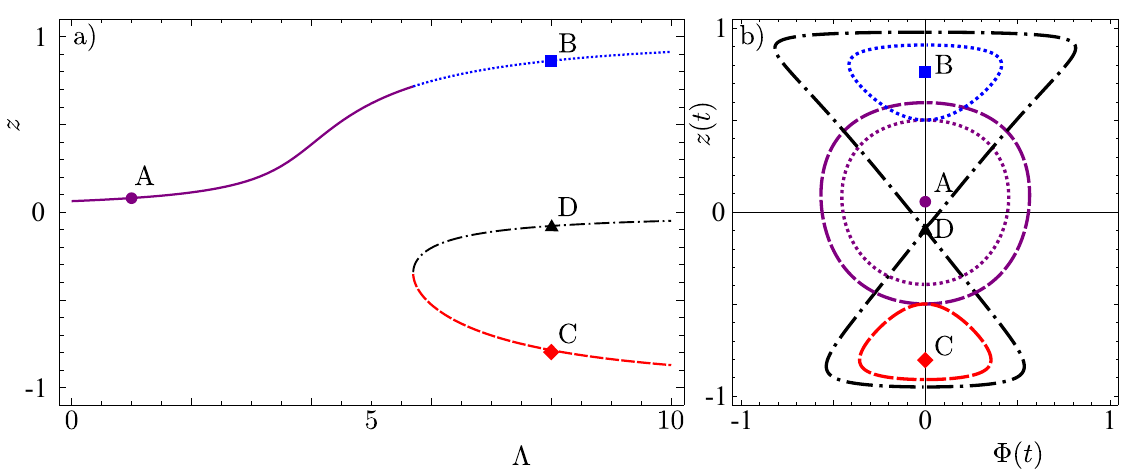}
\caption{ (Color online) Fixed points and typical trajectories for parameter set $\{\Delta = -0.5, \Lambda=1$ and $8, k=10\}$. a) Fixed points for the equations of motion in the weak regime, described by Eqs. (\ref{eq:WLEqMz} -\ref{eq:WLEqMPhi}), at $\Phi = 0$ showing the splitting of a single stable fixed point in the Rabi regime (solid purple) into two fixed points in the Josephson regime (dotted blue and dashed red). The unstable fixed point, or separatrix, in the Josephson regime (dash-dotted black) is also shown. b) Examples of trajectories with symmetrical initial conditions $\{z_{\pm}=\pm 0.5, \Phi=0\}$, leading to asymmetrical Rabi oscillations (purple dotted for $z_{+}$ and dashed for $z_{-}$) around the fixed point A; and asymmetrically localized Josephson oscillations around the fixed point B (blue dotted for $z_{+}$) and C (red dashed for $z_{+}$). Two trajectories starting slightly above and below the separatrix D are also shown (dash-dotted black).}
\label{fig:Fig4}       
\end{figure}

Figure \ref{fig:Fig4} shows an example of asymmetric Rabi and Josephson oscillations brought by the quantum driving field. 
A parameter set $\{ \Delta = 0.5, k=10 \}$ is taken and the fixed points found, Fig.~\ref{fig:Fig4}(a). 
From the fixed points, two coupling ratios are chosen in the Rabi, $\Lambda = 1$, and Josephson, $\Lambda=8$, regimes; this procedure delivers one stable fixed point, $\{ A \}$, in the Rabi regime and two stable, ${B,C}$, and one unstable, $\{ D \}$, fixed points in the Josephson regime. 
Figure \ref{fig:Fig4}(b) shows typical trajectories in these two regimes, where it is possible to see that symmetric initial conditions, \textit{i.e.}, $\{ z_{\pm}(t=0)= \pm z_{0}, \Phi(t=0) = 0 \} $, do not deliver symmetric trajectories.
The latter can be seen straightforward from the position of the unstable fixed point, also called separatrix, and trajectories starting close to it.

\begin{figure}
\includegraphics[scale=1]{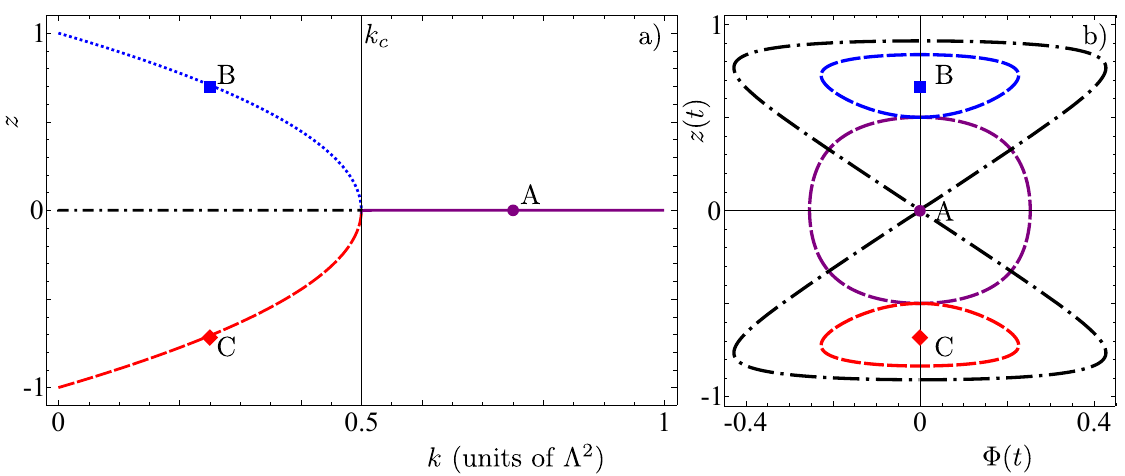}
\caption{ (Color online) Off resonance, $\delta \neq 0$, with $\Delta = \left( \delta + \kappa \right) / \lambda = 0$, \textit{i.e.} $\delta = - \kappa$,  a) Fixed points for the equations of motion in the weak regime, described by Eqs.(\ref{eq:WLEqMz}-\ref{eq:WLEqMPhi}), at $\Phi = 0$ in the large couplings ratio limit, $\Lambda = \xi/\lambda \rightarrow \infty$ (numerical value $\Lambda=1000$), showing the bifurcation of a single stable fixed point in the Rabi regime (solid purple) into two fixed points in the Josephson regime (dotted blue and dashed red). The unstable fixed point, or separatrix, in the Josephson regime (dash-dotted black) is also shown. b) Examples of trajectories with symmetrical initial conditions $\{z_{\pm}=\pm 0.5, \Phi=0\}$, identical Rabi oscillations (purple solid) around the fixed point A (numerical value $k=750000$) and symmetrically localized Josephson oscillations around the fixed point B (blue dotted) and C (red dashed) Two trajectories starting slightly above and below the separatrix D are also shown (dash-dotted black) (numerical value $k=250000$). }
\label{fig:Fig5}       
\end{figure}

In the limit where the couplings ratio is large, $\Lambda \rightarrow \infty$, it is possible to see a pitchfork bifurcations depending on the excitation ratio; an approximate critical excitation ratio can be calculated as $k_{c} \approx \Lambda^2/2$.
At this critical value, a stable fixed point bifurcates into two new stable fixed points and the original fixed point becomes an unstable fixed point acting as a separatrix in phase space. 
There is a symmetry breaking in the dynamics due to the bifurcation, a transition from Rabi to Josephson oscillations; \textit{i.e.}, two initial symmetric states share identical dynamics in the Rabi regime, while in the Josephson regime they localize in different regions of phase space.
Figure \ref{fig:Fig5} shows an example of this symmetry breaking in the dynamics, in which  for large coupling ratios, $\Lambda \gg 1$, \textit{i.e.} $\eta \gg \lambda$, it is possible to locate a pitchfork bifurcation point, $k_{c+} \approx \Lambda^2/2$ even for the off-resonance condition,  $\delta \neq 0$.
Now the transition between the two-level system and the field $\delta$ is different from zero but balances off the nonlinearity, {\textit i.e.}, $\Delta = 0$.
This condition, $\eta \gg \lambda$, relates to the phase space region where maximal shared bipartite  concurrence in the atomic ensemble may be obtained in the quantum treatment of this model \cite{RodriguezLara2010}. The difference comes from the large excitation parameter ratio arising in this semi-classical analysis, $k_{c+} \gg 1$, \textit{i.e.}, $N_{q} \ll n$ as $z_{c} \approx 3 / \Lambda^2 \ll 1$.

\section{Conclusion}
\label{sec:4}

In summary, we have presented an analysis of the classical dynamics of a two-species BEC large in size driven by a quantized field in the presence of nonlinear processes.
In the weak regime, we find that the nonlinear phase from the Kerr nonlinearity, $\kappa$, shifts the  transition detuning of the effective Hamiltonian of a generalized Dicke model; while the $\chi^{(2)}$ nonlinear coefficient re-scales the coupling ratio.
This mean-field Hamiltonian is equivalent to a nonrigid, nonlinear pendulum, for which a transition from Rabi to Josephson dynamics is identified depending on both the intra-BEC interactions to field-ensemble coupling ratio and the ratio between the total excitation number and the ensemble size.
Moreover, we find that the symmetry of Josephson dynamics is broken by the quantum field, and an actual pitchfork bifurcation point is found in the regime where the intra-ensemble interaction is larger than the field-ensemble coupling.
Furthermore, It is known that symmetry-breaking in the classical dynamics of  BEC may herald entangled quantum states \cite{Micheli, Hines}. 
Our results may provide a deeper understanding about the collective dynamics of interacting BECs and give another example in favour of the aforementioned conjecture relating classical and quantum regimes for nonlinear systems.

\begin{acknowledgement}
B.M.R.L. is grateful for the hospitality and camaraderie of the Theoretical Optics Group at National Tsing Hua University, Taiwan.
\end{acknowledgement}

\end{document}